\documentclass{jpp}
\usepackage{graphicx}

\usepackage[utf8]{inputenc}
\usepackage[T1]{fontenc}
\usepackage{amsmath}
\usepackage{soul}
\usepackage{url}
\usepackage{color}

\shorttitle{Unsupervised classification of plasmoid instability}
\shortauthor{S. K\"ohne, E. Boella, M. E. Innocenti}

\title{Unsupervised classification of fully kinetic simulations of plasmoid instability using Self-Organizing Maps (SOMs)}

\author{Sophia K\"ohne\aff{1}, Elisabetta Boella\aff{2,3}, Maria Elena Innocenti\aff{1}\corresp{\email{mariaelena.innocenti@rub.de}}}

\affiliation{\aff{1}Institut f{\"u}r Theoretische Physik, Ruhr-Universit{\"a}t Bochum, Universit{\"a}tstra{\ss}e 150, 44801 Bochum, Germany
\aff{2}Physics Department, Lancaster University, Bailrigg, Lancaster LA11NN, UK
\aff{3}Cockcroft Institute, Sci-Tech Daresbury, Warrington WA44AD, UK}

\begin{document}

\maketitle

\begin{abstract}
The growing amount of data produced by simulations and observations of space physics processes encourages the use of methods rooted in Machine Learning for data analysis and physical discovery. We apply a clustering method based on Self-Organizing Maps \textcolor{black}{(SOM)} to fully kinetic simulations of plasmoid instability, with the aim of assessing its suitability as a reliable analysis tool for both simulated and observed data. We obtain clusters that map well, a posteriori, to our knowledge of the process: the clusters clearly identify the inflow region, the inner plasmoid region, the separatrices, and regions associated with plasmoid merging. SOM-specific analysis tools, such as feature maps and Unified Distance Matrix, provide one with valuable insights into both the physics at work and specific spatial regions of interest. The method appears as a promising option for the analysis of data, both from simulations and from observations, and could also potentially be used to trigger the switch to different simulation models or resolution in coupled codes for space simulations.
\end{abstract}

\section{Introduction}
In recent years, space physics research based on observations and numerical experiments has converged under several points of view. 
First, both spacecrafts and numerical simulations have nowadays easy access to kinetic-scale processes. Recent missions, such as the Magnetospheric MultiScale MMS \citep{burch2016magnetospheric} spacecrafts, Parker Solar Probe PSP \citep{fox2016solar}, and Solar Orbiter \citep{muller2020solar}, routinely sample ion and electrons scales. Similarly, fully kinetic codes are employed for the study of ion and electron scale processes embedded in increasingly large spatial and temporal domains. This is the case in particular for semi-implicit Particle-In-Cell, PIC~\citep{hockney2021computer}, codes, e.g. \citet{markidis2010multi, Innocenti2017b, Lapenta-JPP-2017}, where advanced numerical techniques may also be used to enable the simulation of large domains, either within the fully kinetic description, e.g. \citet{innocenti2013multi, innocenti2015introduction, Innocenti2019a}, or by coupling the kinetic and fluid description, e.g. \citet{daldorff2014two, ashour2015multiscale, lautenbach2018multiphysics}. 
Second, both spacecraft observations and numerical simulations produce an increasing amount of data. As an example, MMS collects a combined volume of $\sim$ 100 gigabits per day of particle and field data, of which only a fraction can be transmitted to the ground due to downlink limitations \citep{baker2016magnetospheric}. At the same time, a single PIC simulation can produce $\sim$ tens of terabytes of output, depending on the frequency at which field and particle information is saved. Such a huge amount of data leads naturally to the usage of Machine Learning (ML, \citep{bishop2006pattern, goodfellow2016deep})-based methods for data analysis and investigation. A review of recent applications of ML techniques to the space physics domain is found in \citet{camporeale2019challenge}. Particularly useful in this context are classification or clustering techniques that can be used for the identification of "similar" regions, either sampled or simulated, and hence for the detection of boundary crossing. Recently, supervised ML techniques have been used for the classification of large-scale magnetospheric regions and detection of boundary crossing, chiefly bow-shock and magnetopause crossings, e.g. in \citet{da2020automatic, argall2020, Breuillard2020, olshevsky2021automated, nguyen2022massive, Lalti2022}. ML techniques, either supervised, e.g. \citet{camporeale2017classification, Li2020}, or unsupervised, e.g. \citet{roberts2020objectively, amaya2020visualizing}, have been used for the classification/ clustering of solar wind states. Unsupervised techniques, namely Gaussian mixture models, have recently proven quite effective in PIC simulations, either for the identification of regions of interest \citep{dupuis2020characterizing}, where the particle distribution functions deviate from Maxwellian, or to encode particle information for later resampling during simulation restarts \citep{chen2021unsupervised}. 
In this paper we will use an unsupervised clustering technique based on Self Organizing Maps (SOM, \citep{kohonen1982self}) to cluster simulated data points obtained from a PIC simulation. The aim is to verify whether this procedure can be used for two purposes, namely simulation pre-processing and  scientific investigation. 
The same procedure (with minimal variations) has already given satisfactory results on data of rather different origin, obtained from both observations and simulations. In~\citet{amaya2020visualizing}, it has been applied to 14 years of Advanced Composition Explorer (ACE,~\citep{stone1998advanced}) solar wind measurements. In~\citet{innocenti2021}, it has been used to cluster simulated data, and specifically data from a global magnetospheric simulation. The code used there was the Magneto-Hydro-Dynamic (MHD)-based code OpenGGCM-CTIM-RCM~\citep{Raeder2003}, which targets large scale processes originating from the interaction of the solar wind with the magnetosphere–ionosphere–thermosphere system. Quite satisfactorily, the clustering procedure used there was able to cluster simulated points into regions associated, a posteriori, with the pristine solar wind, the magnetosheath (divided in three clusters, just downstream and further away from the bow shock), the lobes, the inner magnetosphere and boundary layers. Verification and validation activities were conducted to ascertain the dependence of the obtained clusters from several hyper-parameters, including the features used for the clustering, the number of k-means clusters used for the classification of the trained weights, and SOM-related hyper-parameters, such as the number of nodes in the SOM, the learning rate, and the initial lattice neighborhood width (see Section \ref{sec:SOM}). Robustness of the proposed clustering to temporal variation was also investigated.\\
One rather fundamental question left open in~\citet{innocenti2021} was whether a similar clustering procedure would produce equally meaningful results when applied to smaller scale processes, kinetic in nature. MHD simulations intend to reproduce plasma behavior at scales large enough that certain assumptions can be considered satisfied. Among these assumptions, we list quasi-neutrality, the presence of thermal (Maxwellian) and isotropic velocity distribution functions, the possibility of ignoring the non ideal terms in the Ohm's law and finite gyroradii effects~\citep{ledvina2008modeling}. These assumptions are obviously not respected in most heliospheric plasmas, especially at the very small and fast scales sampled by recent magnetospheric and solar wind missions, such as the above-mentioned MMS, PSP, and Solar Orbiter. Hence, the need arises to verify if the clustering procedure described in ~\citet{innocenti2021} is robust to using simulation approaches, such as PIC methods, that deliver results directly comparable with observations \citep{innocenti2016study}. This work intends to address this question. The aim of this work is two-fold: on one hand, we intend to verify if this procedure is useful in the post-processing of large-throughput numerical simulations. On the other hand, this analysis constitutes a necessary first step to validate the method before applying it to spacecraft observations.
We apply the clustering procedure from~\citet{innocenti2021}  to a fully kinetic, PIC simulation of the plasmoid instability. The plasmoid instability is a fast instability that breaks down current sheets into multiple magnetic islands, which later undergo non-linear evolution. We refer the reader to~\citet{loureiro2015magnetic, pucci2020onset} for a review of recent developments in plasmoid instability research both in the collisional and collisionless regimes. The plasmoid instability results in fast, spontaneous magnetic reconnection in plasmas and, as such, is deeply connected with the fundamental topic of particle heating and acceleration in space and astrophysical plasmas. A number of recent PIC simulations have focused specifically on the role of magnetic reconnection triggered by plasmoid instability in electron heating and acceleration. Processes observed in plasmoid instability simulations that result in particle heating and acceleration are acceleration by the reconnection electric field \citep{li2017particle}, Fermi acceleration \citep{guo2010particle} and plasmoid merging \citep{drake2006electron, Petropoulou2018}. The efficiency of these processes has been observed to vary according to the plasma beta~\citep{li2015nonthermally} and magnetization~\citep{guo2016efficient}. In simulations of plasmoid instability evolution, different regions are immediately distinguishable with the naked eye: an inflow region, separatrices, the plasmoid themselves and the plasmoid merging regions. We aim at understanding if our unsupervised clustering method is capable of comparable region identification.

This paper is organized as follows: in Section~\ref{sec:SOM} we describe SOMs and our clustering procedure, in Section~\ref{sec:PIC} the simulation used. In Section~\ref{sec:res} we describe our results: preliminary data inspection, scaling experiments, SOM training and the k-means clustering of trained SOM nodes (Section~\ref{sec:res}). An a posteriori analysis of clustering results and physical insights obtained from them are described in Section~\ref{sec:posteriori}. Discussions and conclusions follow. In \ref{sec:appendix} we comment on the robustness to hyper-parameter choice of our clustering procedure.

\section{Self Organizing Maps: a summary}
\label{sec:SOM}

Self Organizing Maps, SOM \citep{kohonen1982self, villmann2006magnification}, can be viewed as both a clustering and a dimensionality reduction procedure~\citep{kohonen2014matlab}. The aim is to represent a large set of possibly high-dimensional data as a (usually) two-dimensional (2D) \textit{ordered} lattice composed of $y \times x= q$ nodes/ units/ neurons, with $y$ and $x$ respectively the number of rows and columns, and $q$ the total number of nodes. Each node $i$ is characterized by a position in the 2D lattice and a weight $\boldsymbol{w}_i \in \mathbb{R}^n$, where $n$ is the number of features associated with each data point, and hence with each weight. Before the training, the weights are initialized randomly or as to span the parameter space covered by the first two Principal Components~\citep{shlens2014tutorial} of the input data, to make training faster~\citep{kohonen2014matlab}. At the end of the training, the node weights will have been modified so that the nodes a) represent local averages of the input data that maps to them, and b)  are topographically ordered according to their similarity relation. Nearby nodes should be more "similar" to each other than far away nodes. To define similarity a distance metric is needed. Common choices are Euclidean, sum of squares, Manhattan, Tanimoto distances ~\citep{XuTian2015}.  \\
The training is unsupervised, meaning that training data are not labelled. Each data point $\boldsymbol{x}_\tau$ is presented to the map multiple times. On each occasion the following procedure, a mix of competition and collaboration, is repeated:

\begin{itemize}
    \item competition: the Best Matching Unit, BMU, of the input data $\boldsymbol{x}_\tau$ is identified, by selecting the node whose weight $\boldsymbol{w}_s$, \textcolor{black}{among the set $\boldsymbol{W}$ of all the weights in the map}, has minimum distance ($||.||$) with respect to $\boldsymbol{x}_\tau$ :
    
    \begin{equation}
    \boldsymbol{w}_s = \underset{\boldsymbol{w}_i \in \boldsymbol{W}}{\arg \min}\left( \lVert \boldsymbol{x}_\tau - \boldsymbol{w}_i\rVert \right) 
    \label{eq:winner}
\end{equation}

 \item collaboration: to obtain an ordered map, not only the BMU but also some neighboring nodes are tagged for update at each iteration. Such neighbors are selected through a lattice neighborhood function $h_\sigma(\tau,i,s)$:
 
 \begin{equation}
    h_\sigma(\tau,i,s) = e^{-\frac{\lVert \boldsymbol{p}_i - \boldsymbol{p}_s \rVert^2}{2\sigma(\tau)^2}},
\end{equation}
 
 where $s$ is the BMU index and $\boldsymbol{p}_i \in\mathbb{R}^2 $ the position of node $i$ in the 2D map. $\sigma(\tau)$ is the iteration($\tau$)-dependent lattice neighborhood width, that determines the extent around the BMU of the update introduced by the new input.
 
 \item weight update: the weights of the selected nodes are updated, to make them become more similar to the input data. The magnitude of the correction $\Delta\boldsymbol{w}_i$ for the node $i$ is calculated as follows:
 \begin{equation}
    \Delta\boldsymbol{w}_i = \epsilon(\tau)h_\sigma(\tau_i,i,s)(\boldsymbol{x}_\tau-\boldsymbol{w}_i)
    \label{eq:update}
\end{equation}
The correction depends on the distance of the nodes from the BMU and on the learning rate $\epsilon(\tau)$, which is also iteration-dependent. 
    
\end{itemize}

Both the lattice neighborhood width and the learning rate decrease with increasing iteration number, according to predetermined rules. The initial lattice neighbor width and learning rate, $\sigma(\tau=0)$ and $\eta(\tau=0)$, are labelled $\sigma_0$ and $\eta_0$ respectively. The rationale for this iteration-dependent decrease is the following: at the beginning of the training far-reaching, large-magnitude updates of the node weights are needed, since the node weights are very different from the input data and the topology of the data has to be enforced on the map. After several iterations, when the map already resembles the data, the node updates become targeted in position and smaller in magnitude. The overall convergence of the map is, therefore, separable into two phases: first the topographic ordering of the nodes and then the convergence of the weight values according to the quantization error~\citep{VanHulle2012}.

The quantization error $Q_E$, defined as
\begin{equation}
    Q_E = \frac{1}{m} \sum_{i=1}^m \lVert \boldsymbol{x}_i - \boldsymbol{w}_{s|\boldsymbol{x}_i} \rVert
    \label{eq:Qe}
\end{equation}
measures the average distance between each of the $m$ entry data points $\boldsymbol{x}_i$ and their BMU, $\boldsymbol{w}_{s|\boldsymbol{x}_i} $. As such, it can be used to measure how closely the map reflects the training data distribution. The quantization error is expected to decrease and finally plateau during map training. 

SOMs are at the core of the clustering procedure we use in this work, which has already been used (with minimal variation) and described in~\citet{amaya2020visualizing} and~\citet{innocenti2021}. We briefly describe the method here for the reader's convenience:

\begin{itemize}
\item preliminary data inspection and data pre-processing: we describe in Section \ref{sec:res} our experiments with different data scalers

\item SOM training, as described above

\item k-means clustering~\citep{lloyd1982least} of the trained SOM nodes. After SOM training, the data points are clustered in $q$ clusters, with $q$ the number of SOM node. $q$ is typically too high for meaningful results inspections. Hence, we further cluster the trained SOM nodes in a lower number of k-means clusters. The number of clusters is determined through an unsupervised procedure, the Kneedle cluster number determination~\citep{Satopaa2011} 

\item clustering of the simulated data points, based on the k-means cluster of their BMU. At this stage, we can inspect our clustering results and use the result of the clustering to obtain physical insights on our data, see Section \ref{sec:res} and \ref{sec:posteriori}.
\end{itemize}

In~\citet{amaya2020visualizing} and~\citet{innocenti2021} the serial SOM implementation from~\citet{vettigliminisom} was used. Here, due to the large volume of data points to cluster, we move to the parallel SOM implementation using CUDA \citep{cuda} and C++ by \citet{CUDASOM}. The algorithm is the same as in the serial implementation, but the distance calculations between the data samples and the neuron weights are parallelized. The parallel implementation allows one to choose, for the training phase, between online and batch learning. In the online version of the algorithm, the data samples are each processed according to the scheme described above. 
In this case, the updates of the weights are determined by the impact of the individual samples. The batch algorithm on the other hand, finds the BMU for each sample and then sums up all data samples mapping to one node at the beginning of a training cycle to form node-sums. Then, during training, a neighborhood set for each of the nodes is found and all the nodes-sums are summed up to a neighborhood-sum, which is divided by the total number of samples contributing, to form a neighborhood-mean. In the end the weights are updated in one operation over all nodes of the SOM, where the values of the weights are replaced by those neighborhood-means ~\citep{kohonen2014matlab}. The batch algorithm does not need a learning rate. We chose here the online training algorithm. 
In Table \ref{tab:cudaserial} we compare the average run time per epoch, in seconds, of the serial and parallel SOM implementation, as a function of the number of samples $m$. One epoch corresponds to presenting all the samples to the map once. In the last column we record the ratio of the execution times of the parallel and serial implementation as a function of the number of samples. We observe that using the parallel implementation becomes increasingly convenient with increasing sample number. 

\begin{table}
    \caption{Comparison of the execution times of the parallel CUDA/C++ implementation CUDA-SOM \citep{CUDASOM} and the serial Python implementation miniSom \citep{vettigliminisom}, both trained for 5 epochs with different sample sizes $m$. The data points are extracted from the upper current sheet of the simulation described in Section \ref{sec:PIC}. The SOM parameters used for the training are initial learning rate $\eta_0=0.5$ and  initial neighborhood radius $\sigma_0=0.2 \times\max(x,y)$.}
    \centering
    \begin{tabular}{c c c c}
    Implementation & $m$ & average runtime/epoch in s & speedup factor\\
    \hline
    serial & 2264 & 0.785 & \\
    parallel &  & 0.314 & 2.4\\
    \hline
    serial & 5660 & 2.476 & \\
    parallel &  & 0.798 & 3.1\\
    \hline
    serial & 11320 & 2.986 & \\
    parallel &  & 1.603 & 3.7\\
    \hline
    serial & 22641 & 14.625 & \\
    parallel &  & 3.240 & 4.5\\
    \hline
    serial & 226419 & 356.094 & \\
    parallel &  & 36.302 & 9.8\\
    \hline
    serial & 1132096 & 3597.720 & \\
    parallel &  & 176.865 & 20.3\\
    \end{tabular}
    \label{tab:cudaserial}
\end{table}

\section{Simulation description}
\label{sec:PIC}

\begin{figure}
	\centering
    \includegraphics[width=\textwidth]{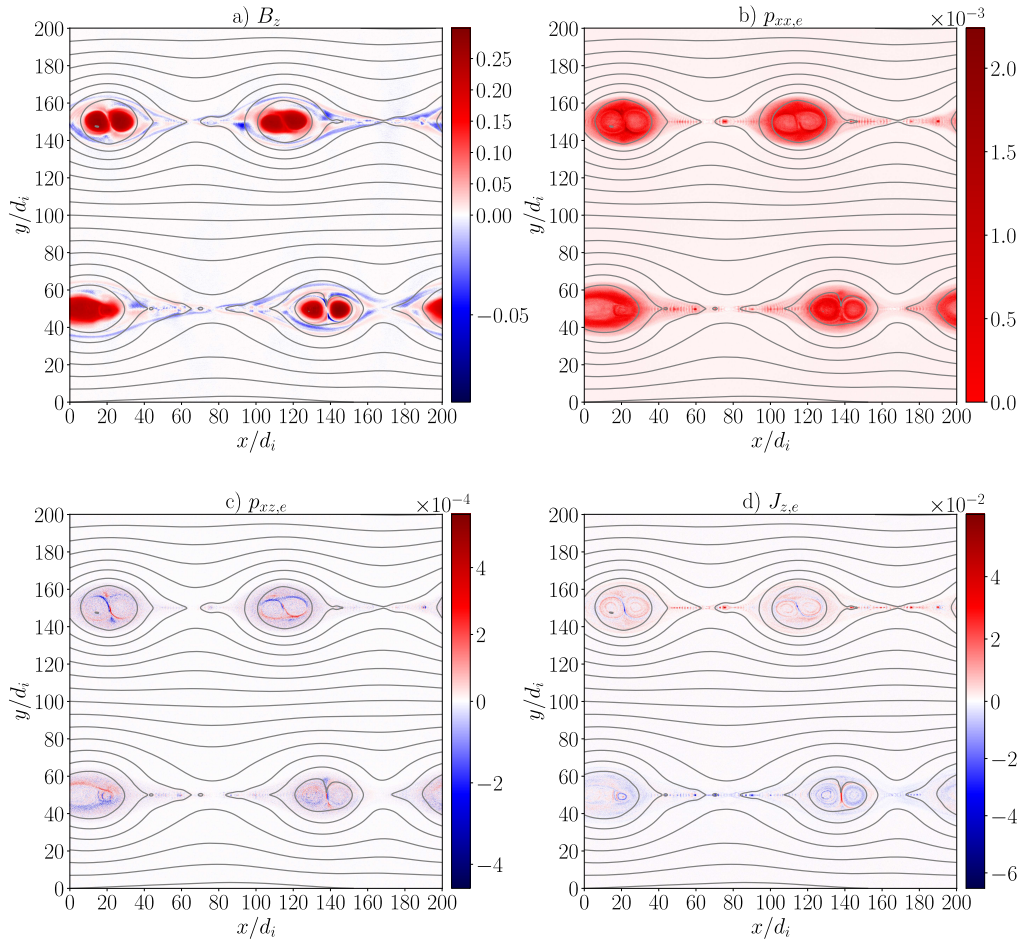}
        \caption{Out-of-plane magnetic field component ($B_z$, panel a), one diagonal ($p_{xx,e}$, panel b) and one non-diagonal, ($p_{xz,e}$, panel c), electron pressure term, and the out-of-plane electron current ($J_{z,e}$, panel d) at $\Omega_{ci}t=320$. Magnetic field lines are superimposed in black. Electromagnetic fields, pressure components and currents are in units of $m_i c \omega_{pi} / e$, $n_0 m_i c^2$  and $e n_0 c$, respectively.}
        \label{fig:raw:fieldlines}
\end{figure}

Self Organizing Maps are used to cluster results of a 2D kinetic simulation of plasmoid instability. The simulation was carried out with the semi-implicit PIC code ECsim \citep{Lapenta-JPP-2017,Lapenta-JCP-2017,Gonzalez-CPP-2018}. In the PIC algorithm, a statistical approach is adopted and plasmas are described using computational particles or macroparticles representative of several real plasma particles. These computational particles interact via the electromagnetic fields that they themselves produce. The fields are computed on a fixed grid by solving Maxwell's equations. ECsim implements a semi-implicit algorithm, which makes the code stable and accurate over different spatial and temporal resolutions. As a consequence, the code resolution can be tuned to the physics of interest, rather than to the smallest scales of the problem \citep{Micera-ApJ-2020}.
In the simulation, two oppositely directed force-free current sheets are used as initial condition. The sheets sustain a magnetic field $\boldsymbol{B}= B_x(y) \hat{\boldsymbol{x}} + B_z(y) \hat{\boldsymbol{z}}$ with
\begin{equation}
B_x(y) = B_{0,x} \left[-1 + \tanh \left(\frac{y-0.25 L_y}{\delta} \right) + \tanh \left(-\frac{y-0.75 L_y}{\delta} \right)  \right]
\end{equation} and 
\begin{equation}
B_z(y)=B_{0,x}  \sqrt{ \operatorname{sech}^2 \left(\frac{y-0.25 L_y}{\delta} \right) + \operatorname{sech}^2 \left(\frac{y-0.75 L_y}{\delta} \right) + B_g^2 }.
\end{equation}
Here, $L_y$ is the transverse size of the simulation box, $\delta = d_i$ is the current sheet half thickness, $d_i = c/\omega_{pi}$ is the ion inertial length, $\omega_{pi} = \sqrt{4 \pi e^2 n_{0i}/m_i}$ the plasma frequency for ions of density $n_{0i}$ and mass $m_i$, $e$ is the elementary charge and $B_g$ the magnitude of the guide field. While the upper current sheet is unperturbed and the tearing instability is seeded from numerical noise, the magnetic field of the lower current sheet was perturbed with a long wave perturbation \citep{Birn-JGR-2001} to trigger the instability on faster time scales. The simulation box is filled with a plasma composed of electrons and ions with mass ratio $m_i/m_e = 25$, where $m_e$ is the electron mass, having uniform density $n_{0i} = n_{0e} = n_0$ and temperature $T_{0i} = T_{0e} = T_0$. Electrons are initialised with a drift velocity $\boldsymbol{v}_e = v_{e,x}(y) \hat{\boldsymbol{x}} + v_{e,z}(y) \hat{\boldsymbol{z}}$, such that $\nabla \times \boldsymbol{B} = 4 \pi \boldsymbol{J}_e/ c$ is satisfied, with $\boldsymbol{J}_e$ the electron current density. In our simulation, we set $v_A = 0.2 \, c$, $\omega_{pe} / \Omega_{ce} = 1$, $\beta_e = \beta_i = 0.02$ and $B_g = 0.03 \, B_{0,x}$. Here $v_A = B_{0,x} / (4 \pi n_0 m_i)$ is the Alfv\'{e}n velocity, $\omega_{pe} = \sqrt{4 \pi e^2 n_{0}/m_e} $ is the electron plasma frequency, $\Omega_{ce} = e B_{0,x}/(m_e c)$ is the electron cyclotron frequency, $\beta_{e,i} = 8 \pi n_0 T_0/ B_{0,x}^2$ is the ratio between electron/ion pressure and magnetic pressure and $c$ is the speed of light in vacuum. These dimensionless parameters are similar to those used in \citet{li2017particle} and correspond to plasma conditions typical of the solar corona and the accretion disk corona. We used a simulation box with longitudinal and transverse sizes $L_x = L_y = 200 \, d_i$, discretized with $2128 \times 2128$ cells to resolve $c/\omega_{pe}$ twice. To model the plasma dynamics accurately 64 particle per cell per species were employed. Particles were pushed for more than 7000 iterations with a temporal time step of $0.16 \, \Omega_{ci}^{-1}$ with $\Omega_{c,i}$ ion cyclotron frequency. Periodic boundary conditions for fields and particles were adopted. We performed a detailed convergence study to ensure that the chosen numerical parameters do not affect the physics under investigation. 
Fig.~\ref{fig:raw:fieldlines} displays the out-of-plane magnetic field component ($B_z$, panel a), one diagonal ($p_{xx,e}$, panel b) and one non-diagonal, ($p_{xz,e}$, panel c), electron pressure term, and the out-of-plane electron current ($J_{z,e}$, panel d) at $\Omega_{ci}t=320$. Field lines are superimposed on each panel. At this time the tearing instability is in its non-linear phase in both current sheets. In the upper current sheet, we observed the formation of small magnetic islands that grew with time and merged to produce the shown configuration. 
In the lower current sheet, where a perturbation was originally present, single X point reconnection developed initially. Later, plasmoids formed in the current sheet and plasmoid merging was also observed. Signatures associated with plasmoid merging are described e.g. in \citet{cazzola2015electron, cazzola2016electron}.
At the location of the X points ($x/d_i \simeq 70$ and $170$ in the upper current sheet and $x/d_i \simeq 80$ in the lower current sheet), the out-of-plane magnetic field component $B_z$ shows the typical quadrupolar structure associated with collisionless Hall reconnection. The $xx$ component of the electron pressure tensor is low at the X points and high at the periphery of magnetic islands as noted for the electron temperature in \citet{Lu-ApJ-2019}. A similar behaviour is observed for the $xz$ pressure tensor component, which is higher in absolute value at the border of magnetic islands with respect to other regions. It is interesting to notice that $p_{xz,e}$ changes polarity within a magnetic island. The role of the pressure tensor off-diagonal components in causing fast reconnection in collisionless plasmas has been studied in a number of previous works, e.g. \citet{Hesse-JGR-1998, Ricci-PoP-2004}.

\section{Clustering results }
\label{sec:res}

\begin{figure}
	\centering
	\includegraphics[width=0.8\textwidth]{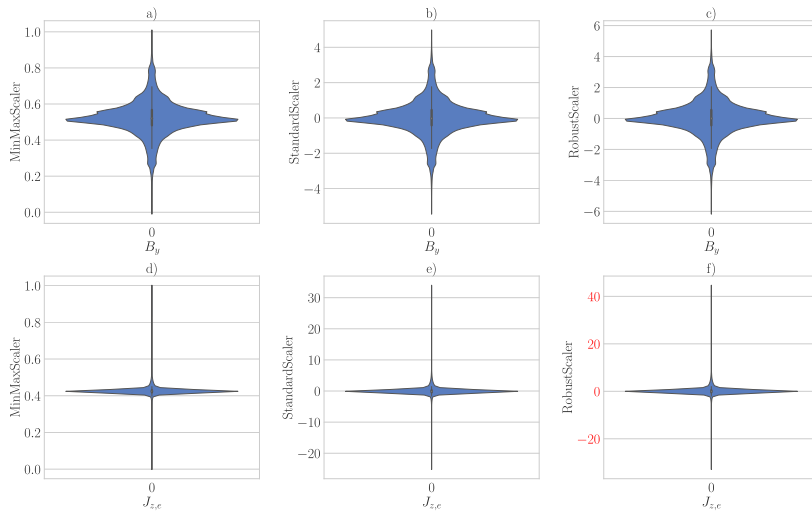}
    \caption{Violin plots of the distribution, after scaling, of an outlier-poor ($B_y$) and an outlier-rich ($J_z,e$) feature. The \textit{MinMax}, \textit{Standard} and \textit{Robust} scalers are used. }
    \label{fig:scalers:effect}
\end{figure}

The data points we cluster are from the upper current sheet of the simulation described in Section \ref{sec:PIC}, at time $\Omega_{ci}t=320.$
As it is standard practice in clustering procedures based on distance, the data are scaled to prevent the features with larger magnitude from dominating the clustering ~\citep{angelis2000}. \textcolor{black}{Scaling the data is the transformation of the feature values according to a defined rule. This is necessary to assure that all features have the same (or rather a proportional) degree of influence in the evaluation ~\citep{huang2015}. If the feature values range for one feature in the tens and another in the thousands a machine learning algorithm will always be influenced more by the feature ranging in the thousands, if scaling is not used.} We tested three different scalers from the scikit-learn library ~\citep{scikit-learn}: the so-called \textit{MinMax}, \textit{Standard} and \textit{Robust} scaler. The first scales each feature in the dataset to a fixed interval, here [0,1], the second to zero mean and unit variance. The third removes the median and scales the data according to a quantile range, here the interquartile range between the 1st and 3rd quartile.
In Fig.~\ref{fig:scalers:effect} we depict the violin plots for the distributions of an outlier-poor feature, $B_y$, and an outlier-rich one, $J_{z,e}$, after scaling them with the three scalers: one can observe that the data range after scaling changes dramatically for the outlier-rich feature. Since $J_{z,e}$ is strongly associated with reconnection X-points, we can expect that the choice of scaler will have an effect on clustering results.

For each set of scaled data, we proceed to train a SOM. The number $q$ of nodes of each map has been determined using the following rule of thumb~\citep{kohonen2014matlab}:
\begin{equation}
q \approx 5\times\sqrt{m}, 
\end{equation}
where $q$ is approximated to an integer and $m$ is the number of samples.
The ratio of the side-lengths $x$, $y$ of the map is set to match the ratio of the largest two principal components, PC,~\citep{shlens2014tutorial} of the training data. We notice that, since the principal component analysis is performed on scaled data, PC, and hence the number of rows and columns in the map, may differ in the different cases. The initial neighborhood radius is then chosen to cover 20 \% of the larger side-length of the map:
\begin{equation}
\sigma_0 = 0.2\times\max{(x,y)}.
\end{equation}
For the \textit{MinMax} and the \textit{Standard} scaler this resulted in $x=y=82$ and $\sigma_0=16$ and for the \textit{Robust} scaler in $x=93$, $y=71$ and $\sigma_0=19$.

\begin{figure}
	\centering
		\includegraphics[width = \textwidth]{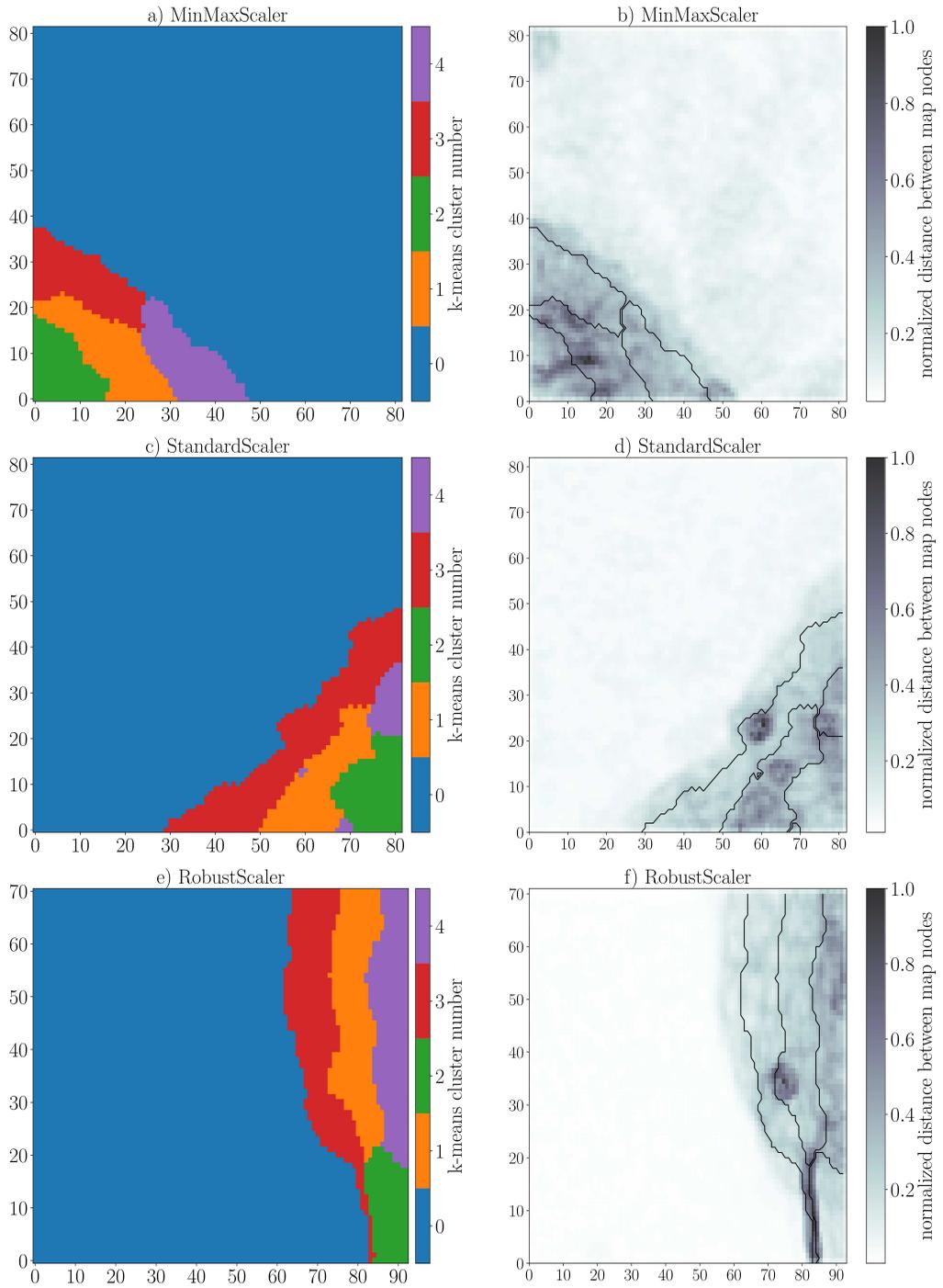}
	    \caption{Trained SOM nodes colored according to their k-means clusters (panel a, c, e) and Unified Distance Matrix maps (panel b, d, and f) for the data scaled with \textit{MinMaxScaler} (panel a and b), \textit{StandardScaler} (panel c and d), \textit{RobustScaler} (panel e and f). Cluster boundaries are drawn in black in panel b, d, and f.}
	    
	    \label{fig:scalers:UDMmmst}
	   
	\end{figure}

\begin{figure}
	\centering
		\includegraphics[width = 0.7\textwidth]{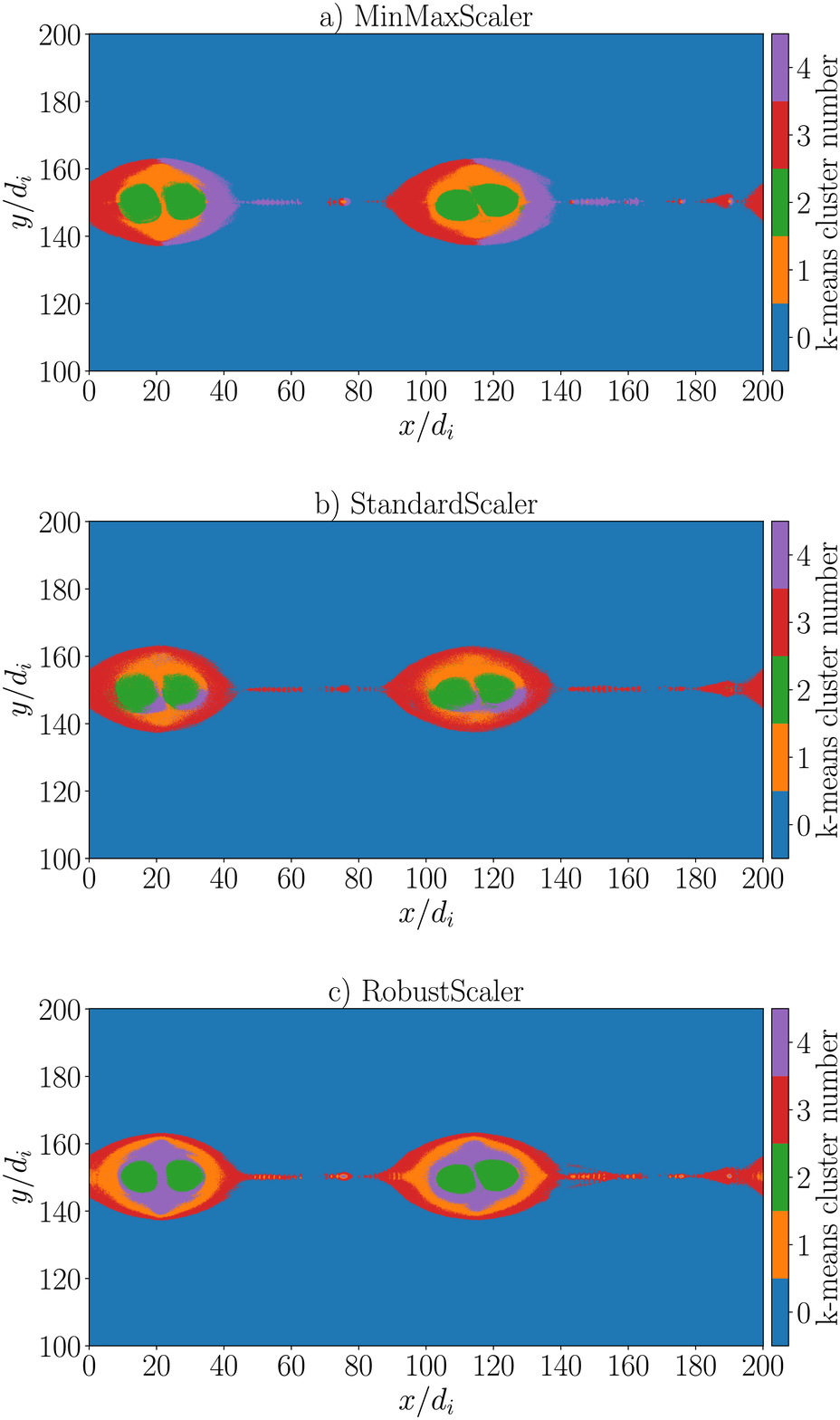}
	    \caption{Upper current sheet simulated data at $\Omega_{ci} t=320$, colored according to the k-means cluster the BMU of each point is clustered into. Data scaled with \textit{MinMaxScaler}, \textit{StandardScaler} and  \textit{RobustScaler} are depicted in panel a, b and c respectively. }
	    
	    \label{fig:scalers:PP}
	   
	\end{figure}

The initial learning rate was kept constant at $\eta_0 = 0.5$ and the weights were initialized \textcolor{black}{using random samples from the training data}.  The maps shown in Fig.~\ref{fig:scalers:UDMmmst} have all been trained for five epochs. We remark that, with this relatively low number of epochs, the map cannot be expected to have converged, as it is often the case with SOMs. However, the results that we obtain are remarkably stable to all parameters, including the number of epochs, as shown in \textcolor{black}{section} \ref{sec:appendix}. There, we study how varying the SOM hyper-parameters \textcolor{black}{and the seed for the random initialization of the weights} influences clustering results. \textcolor{black}{In Table \ref{tab:hyperparams} we report all hyper-parameters used for SOM training.}\\
\begin{table}
     \caption{\textcolor{black}{CUDA-SOM hyperparameters used in SOM training.}}
     \centering
     \begin{tabular}{c c}
     Parameter & Value \\
     \hline
     learningmode & online 'o' \\
     nRows x & 82 (\textit{MinMax}- and \textit{StandardScaler}), 93 (\textit{RobustScaler})\\
     nColumns y & 82 (\textit{MinMax}- and \textit{StandardScaler}), 71 (\textit{RobustScaler})\\
     initial learning rate $\eta_0$ & 0.5\\
     final learning rate $\eta_f$ & 0\\
     initial neighborhood radius $\sigma_0$ &  $0.2\times\max{(x,y)}$\\
     distance function & euclidean 'e'\\
     neighborhood functin & gaussian 'g'\\
     initialization & random input vectors 'c'\\
     lattice & hexagonal\\
     toroidal & off\\
     randomize & on\\
     exponential decay of $\eta$ and $\sigma$ & both 'b'\\
     normalizedistance & off\\   
     \end{tabular}
     \label{tab:hyperparams}
 \end{table}
In Figs. \ref{fig:scalers:UDMmmst} and  \ref{fig:scalers:PP} we show the clustering results obtained with the three scalers. The trained SOM nodes have been further clustered with k-means into larger-scale regions, as described in Sec. \ref{sec:SOM}. The optimal k-means cluster number, $k=5$, has been selected with the Kneedle cluster number determination~\citep{Satopaa2011}. \\
In Fig.~\ref{fig:scalers:UDMmmst}, first column, we see the trained SOM nodes obtained with the three different scalers (\textit{MinMaxScaler}, \textit{StandardScaler} and \textit{RobustScaler}, panel a, c and e, respectively) and colored according to the k-means cluster they are clustered into. In the second column, panel b, d and f, we depict the Unified Distance Matrices, UDM, associated with the three cases. In the UDM representation each neuron is colored according to its normalized distance with respect to its nearest neighbor ~\citep{kohonen2014matlab}: "darker" neurons are less similar to their neighbor than "lighter" neurons. We can observe a similar pattern in the three UDMs: a wide, light area composed of similar neurons is mapped to the same k-means cluster, the 0, blue cluster. A darker area is present in one of the map corners, not necessarily the same one in the three plots because, in SOMs, the information of interest is not the \textit{absolute position}, but the \textit{relative position} with respect to the neighboring nodes. This darker area maps to further four clusters. Even darker areas may be present within clusters or at the boundary between clusters, see e.g. the dark area in Fig. \ref{fig:scalers:UDMmmst} panel f, at the intersection between cluster 3, red, 0, blue and 2, green.\\

In Fig. \ref{fig:scalers:PP} we depict the simulated data points at time $\Omega_{ci}t= 320$, see Fig. \ref{fig:raw:fieldlines} in Section~\ref{sec:PIC}, colored according to the k-means cluster they map to, for the three different scalers. We can now map the neurons from Fig. \ref{fig:scalers:UDMmmst} to the physical regions in Fig.  \ref{fig:scalers:PP}.

In all three cases, the blue cluster from Fig. \ref{fig:scalers:UDMmmst} maps to the plasma outside of the plasmoid region. We can call this region, slightly oversimplifying, the inflow region. The plasmoid region is clustered differently in the three cases: we have here a proof of the importance of pre-processing activities (in this case, the choice of scaler) in clustering results. With \textit{MinMaxScaler}, Fig. \ref{fig:scalers:PP} panel a, the outer plasmoid region neighboring the inflow region is divided into two clusters, cluster 3, red and 4, purple. The inner plasmoid region, readily identifiable "by eye" in the plots of Section \ref{sec:PIC}, is mapped to cluster 2, green. Intermediate plasmoid regions are mapped to cluster 1, orange. With \textit{StandardScaler}, panel b in Fig \ref{fig:scalers:PP}, the outer plasmoid region is assigned to a single cluster, cluster 3, red. The inner plasmoid region is again mapped to cluster 2, green. The remaining plasmoid regions are clustered into two clusters, cluster 1, orange, and 4, purple. The blue and green clusters obtained with the \textit{RobustScaler} are very similar to those obtained with \textit{MinMaxScaler} and \textit{StandardScaler}. "Walking" in  Fig.~\ref{fig:scalers:PP}, panel c, from the inflow region towards the inner plasmoid region we encounter cluster 3, red, 1, orange, and 4, purple. Incidentally, we notice that we can similarly walk from cluster 0 to 3, 1, 4 and finally 2 also in the map depicted in in Fig \ref{fig:scalers:UDMmmst}, panel e, a confirmation that neighborhood in the SOM derives from feature similarity.\\

When confronted with different clustering results, one has to identify criteria that allow one to prefer a clustering method over another. This is usually quite a daunting task for unsupervised methods, where the "ground truth" is not known. Luckily, here, as already in~\citet{innocenti2021}, we are in a rather fortunate situation: we are clustering simulated data, hence we have complete information about all features as a function of space and time. Furthermore, we have previous knowledge of the process we are analyzing. Hence, we can determine that the clustering obtained with the \textit{RobustScaler} is the most useful for our purposes, because it separates regions where we can expect different physical processes to take place. 

It is instructive to speculate on the reasons why regions that we may want to see clustered together, based on the physics that occur there, are assigned to different clusters. This is the case, for example, of cluster 3, red and 4, purple, in Fig.~\ref{fig:scalers:PP} panel a, depicting data scaled with the \textit{MinMaxScaler}. Analyzing the feature values of the points in the two clusters, one realizes that the main difference between the two sets of points is the sign of the $y$ component of the magnetic field. Most other features, including features that we consider of particular relevance in plasmoid instability/ magnetic reconnection simulations (non diagonal pressure terms,  out of plane electron current ...) present quite similar values in the two clusters. Looking at $B_y$ and $J_{z,e}$ after scaling with \textit{MinMaxScaler} in Fig.~\ref{fig:scalers:effect}, panel a and d, we realize that the differences in the outlier-poor $B_y$ for the two clusters are at the two extremes of the value range, here 0 and 1, while the similarities in the outlier-rich $J_{z,e}$ are compressed towards 0.5: \textcolor{black}{the differences in the former "weigh" more in the clustering than the similarities in the latter.} \textcolor{black}{With \textit{RobustScaler}, instead, $B_y$ spans $\sim [-6; 6]$, while $J_{z,e}$ varies between $\sim [-25; 42]$: the $B_y$ values become significantly less relevant in the clustering than the $J_{z,e}$ values, and, according to the results of Fig. \ref{fig:scalers:PP}, are not capable anymore to drive the formation of cluster 3 vs cluster 4.\\ }
We consider this quite a significant demonstration of the importance of accurate pre-processing before clustering activities: data should be inspected before and after scaling, to assess if scaling has preserved important characteristics of the data one may want to rely upon during clustering (e.g., outliers). Furthermore, the results of clustering activities when using different scalers should be compared.

\section{A posteriori analysis of clustering results}
\label{sec:posteriori}
In Section \ref{sec:res} we identified the results obtained using the \textit{RobustScaler} as the most useful for our purposes, since the points clustered together are the ones where we can expect similar processes to occur. This determination is made on the base of our previous knowledge of plasmoid instability evolution.\\

Now, we intend to examine these specific results in more details.\\

First, we focus on the UDM depicted in Fig.~\ref{fig:scalers:UDMmmst}, panel f. The area associated with the inflow region, cluster 0, blue in Fig.~\ref{fig:scalers:PP}, panel c, is lighter in color with respect to the other two cases, showing minimal differences between the nodes. We observe that a darker UDM region encloses a "walled in" area in the lower right corner of the map associated with cluster 2, green. The nodes there are quite similar to each other, as seen from their light color. They are however very different from the ones in the inflow region and also quite different from the other nodes mapping to plasmoids, as seen from the dark color of the nodes at the boundary between clusters. 
Comparing Fig.~\ref{fig:scalers:UDMmmst}, panel f and~\ref{fig:scalers:PP}, panel c, we see that the "walled in" region in the UDM maps to the inner plasmoid region. The features in this regions are, in fact, quite distinct with respect to the neighboring points, as confirmed by violin plot analysis of the data clustered in each cluster and by Fig.     \ref{fig:Brazil} below. Moving from cluster 2 in Fig.~\ref{fig:scalers:PP} towards the inflow region, cluster 0, blue, we encounter cluster 4, purple, 1, orange, and finally cluster 3, red, at the boundary between the inflow region and the plasmoid. We can broadly identify cluster 3, red, as the "separatrix" cluster.

\begin{figure}
	\centering
	\includegraphics[width=1\textwidth]{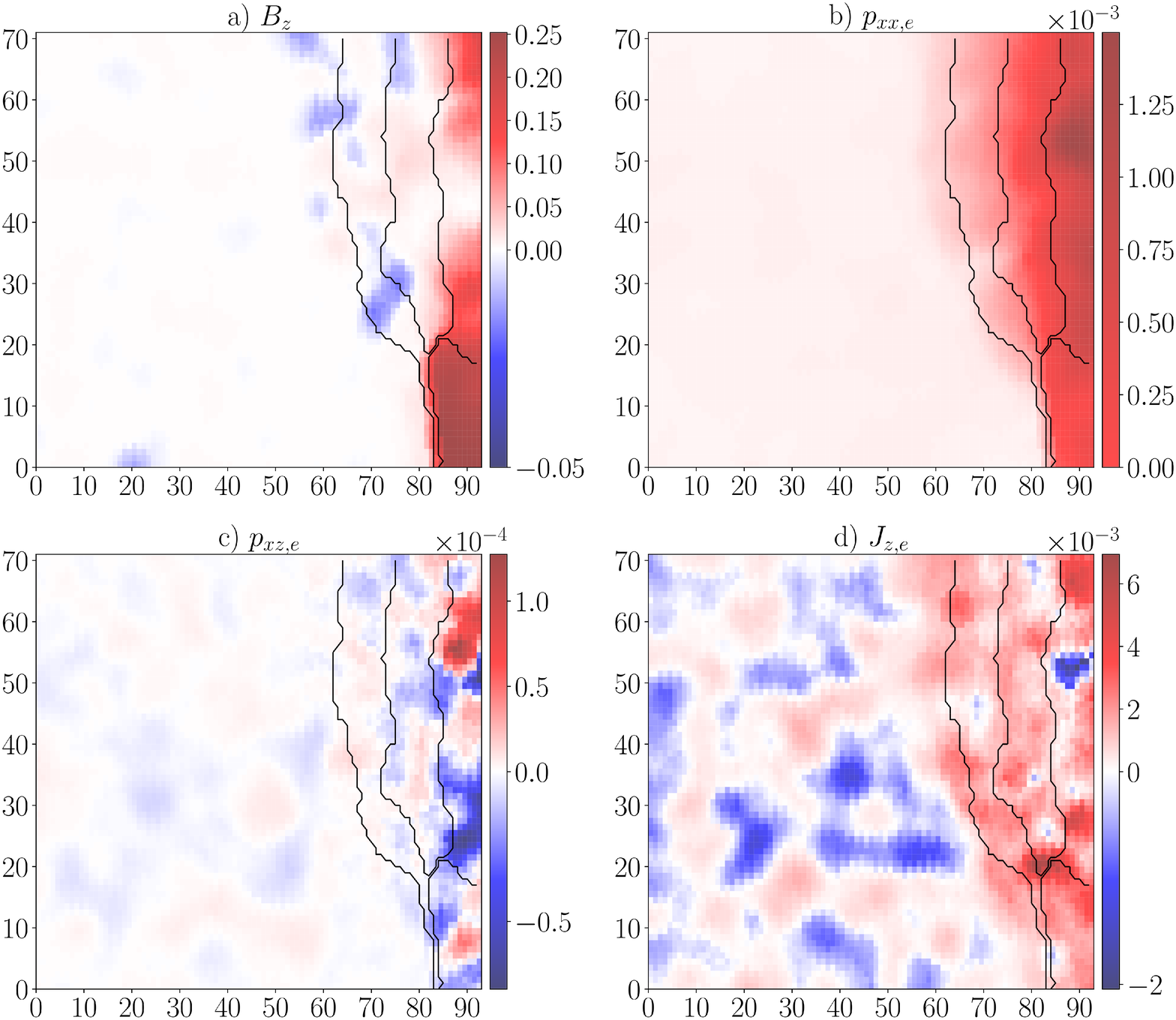}
    \caption{$B_z$ (panel a), $p_{xx,e}$ (panel b), $p_{xz,e}$ (panel c) and $J_{z,e}$ (panel d) values associated with the trained SOM nodes, for data points scaled with the \textit{RobustScaler}. Cluster boundaries are drawn in black.}
    \label{fig:featuremap}
\end{figure}

In Fig.~\ref{fig:featuremap} we depict the variation across the map of $B_z$, $p_{xx,e}$, $p_{xz,e}$, $J_{z,e}$. The boundaries between the k-means clusters are depicted as black lines. The feature values have been scaled back to the normalized values obtained as simulation results. 
\textcolor{black}{Before going into a detailed analysis of Fig.~\ref{fig:featuremap}, we notice several sub-structures in the clusters depicted. This is especially the case for cluster 4, at the right edge of the map. It is therefore no surprise that cluster 4 breaks into 2 and 3 clusters mapping the larger scale sub-structures with $k=6$ and $k=7$ respectively. }

\begin{figure}
	\centering
 	\includegraphics[width=0.6\textwidth]{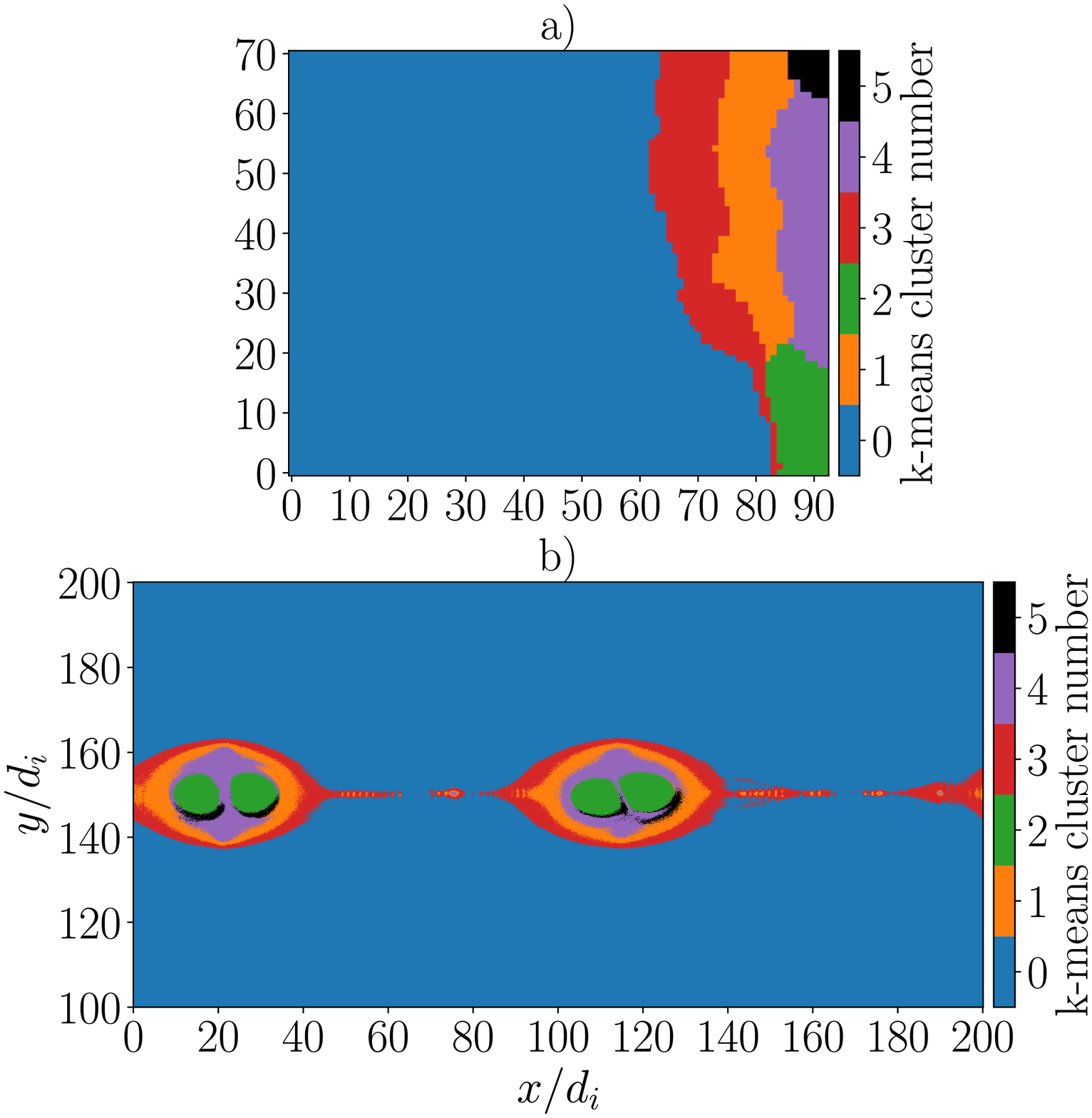}
    \caption{Identification in the simulated plane of regions of interest in the feature maps: the color black is used to highlight nodes of interest in panel a and associated points in the simulation in panel b. }
    \label{fig:handpick_highJz}
\end{figure}

We compare Fig.~\ref{fig:featuremap} with Fig.~\ref{fig:scalers:UDMmmst} and Fig.~\ref{fig:scalers:PP} to highlight important characteristics of the different simulated regions. We observe that high-magnitude, positive values of $B_z$ (Fig. \ref{fig:featuremap}, panel a), are concentrated in the inner plasmoid region, cluster 2, green. Lower, positive values are associated with cluster 4, purple. Most negative $B_z$ values are found in cluster 1, orange and 3, red. In the latter, both positive and negative $B_z$ values are found, most probably associated with the quadrupolar structure in the out of plane magnetic field associated to collisionless reconnection. 
In Fig.~\ref{fig:featuremap}, panel b and c, we depict $p_{xx,e}$ and $p_{xz,e}$. We notice in panel b low $p_{xx,e}$ values in the inflow region, and higher pressure values in correspondence with the plasmoids. 
We further observe that $p_{xx,e}$ exhibits rather high values in the upper right corner of the map, associated with negative values of $p_{xz,e}$ (panel c) and rather high positive value of the out of plane electron current, $J_{z,e}$, depicted in panel d: this interesting region deserves further consideration.
We highlight in Fig.~\ref{fig:handpick_highJz} some of these nodes. The nodes themselves are depicted in black in Fig. \ref{fig:handpick_highJz}, panel a, and the associated points are depicted in the same color in the 2D simulated plane in panel b. We observe that these nodes correspond to very specific regions in cluster 4 characterized by signatures associated with plasmoid merging.

\begin{figure}
	\centering
     \includegraphics[width=0.6\textwidth]{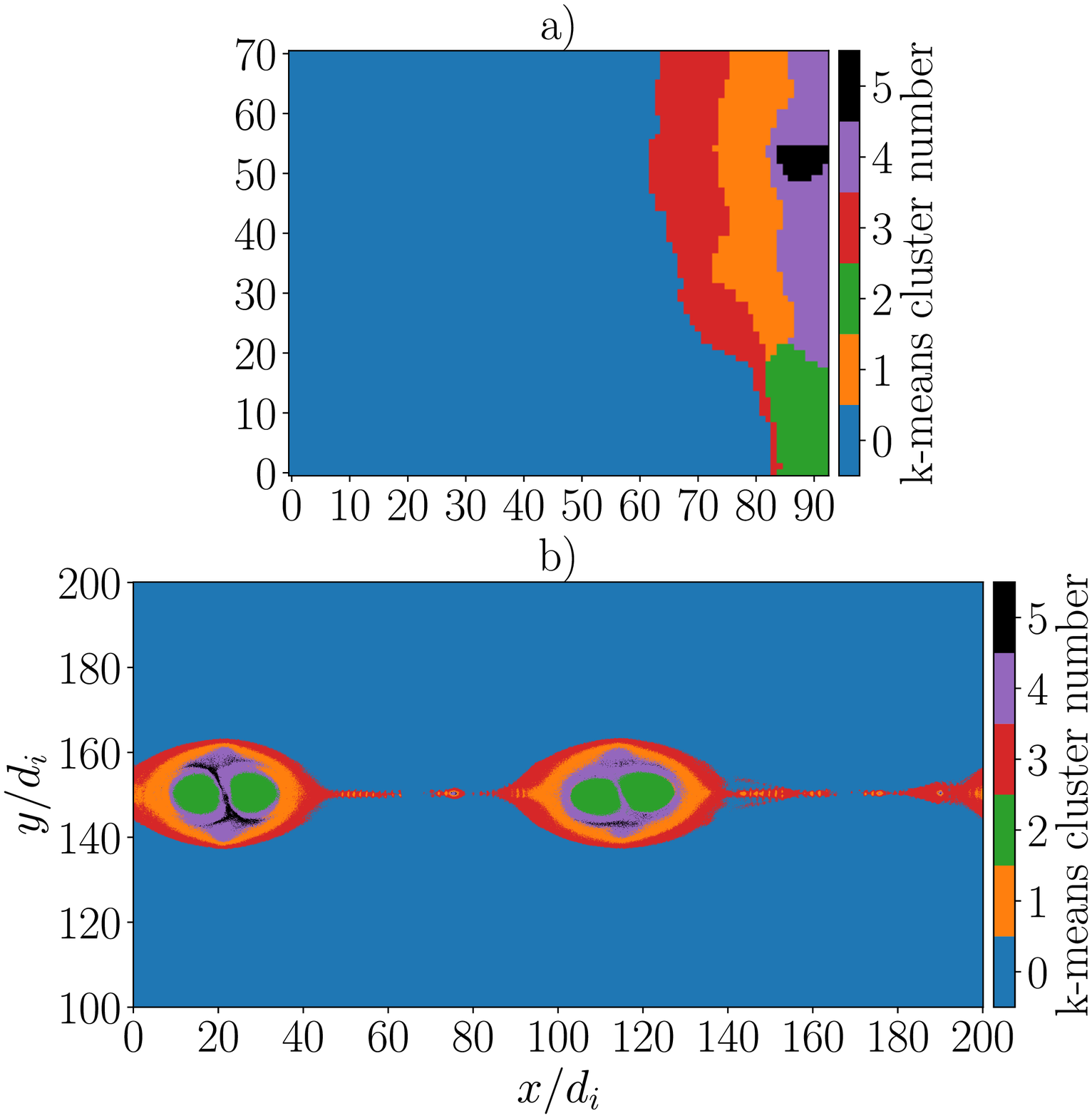}
    \caption{Identification in the simulated plane of regions of interest in the feature maps: the color black is used to highlight nodes of interest in panel a and associated points in the simulation in panel b.}
    \label{fig:handpick_lowJz}
\end{figure}

Going back to Fig.~\ref{fig:featuremap}, we see that some of the nodes below the region just analized are characterized by large positive values of both $p_{xx,e}$ and high, negative values of $J_{z,e}$. Again, in Fig. \ref{fig:handpick_lowJz}, we highlight these nodes in panel a and check with which data points they correspond in panel b. Also in these case, we obtain specific regions in cluster 4, purple, strongly associated to plasmoid merging. We can therefore reaffirm that cluster 4, purple, is the one more directly associated with plasmoid merging, and that specific nodes in the SOMs map to rather localized regions in space where merging-specific signatures are particularly strong.

\begin{figure}
	\centering
     \includegraphics[width=0.6\textwidth]{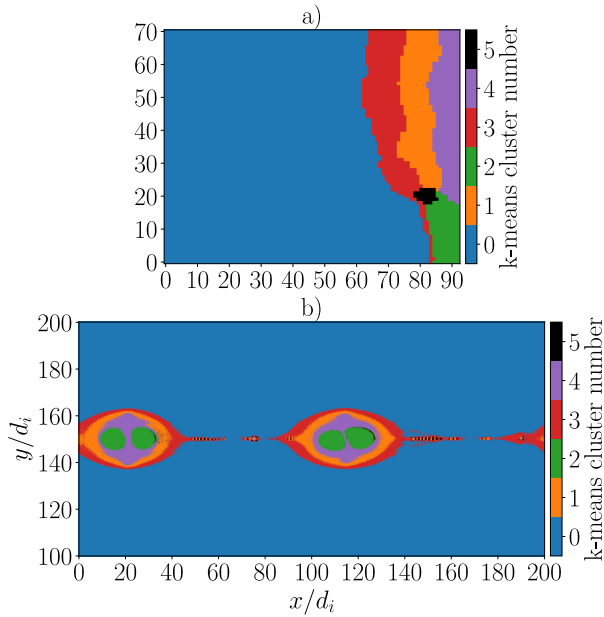}
    \caption{Identification in the simulated plane of regions of interest in the feature maps: the color black is used to highlight nodes of interest in panel a and associated points in the simulation in panel b.}
    \label{fig:handpick_cluster_intersection}
\end{figure}

The variability of $J_{z,e}$ over the map (Fig.~\ref{fig:featuremap}, panel d) is extremely interesting to examine. First of all, we notice that $J_{z,e}$ in the feature map varies over values which seem quite different from those observed in Fig.~\ref{fig:scalers:PP}. This may occur in SOM, and it has already been commented upon e.g. in \citet{innocenti2021}: the feature value associated with a node may be quite different to those of the data points associated with it especially, as in this case, for outlier-rich features. We identify in panel d well specific patterns: for example, positive and negative values of $J_{z,e}$ are associated with the plasmoid merging regions examined in Fig.~\ref{fig:handpick_highJz} and \ref{fig:handpick_lowJz}.
A further pattern of interest in the $J_{z,e}$ feature map is the high-value region at the intersection between cluster 3, 2 and 1. Fig.~\ref{fig:handpick_cluster_intersection} shows that also these nodes map to a rather small region in space, namely smaller-scale plasmoids developing at the X-line, a feature commonly observed in PIC simulations e.g. in \citet{innocenti2015evidence, innocenti2017switch, li2017particle}.

In Fig.~\ref{fig:Brazil} we depict the position of data points as a function of the electron parallel plasma beta $\beta_{\parallel, e}$, $x$ axis, and of the electron perpendicular to parallel temperature ratio, $T_{\perp,e}/ T_{\parallel,e}$, $y$ axis. In panel a we depict all points at initialization, in panel b to f the points associated with cluster 0, 1, 2, 3 and 4 respectively at time $\Omega_{ci}t= 320$. The solid, dashed and dotted lines in the $T_{\perp,e}/ T_{\parallel,e}< 1$ semiquadrant are the isocontours of growth rates $\gamma /\Omega_{ce} = 0.001, 0.1, 0.2$ of the resonant electron firehose instability from~\citet{gary2003resonant}. The solid and dotted lines in the $T_{\perp,e}/ T_{\parallel,e}> 1$ semiquadrant are the  isocontours of growth rates $\gamma /\Omega_{ce} = 0.01, 0.1 $ for the whistler temperature anisotropy instability~\citep{gary1996whistler}. The colors depict the number of points per pixel. This visualization is quite common for both electrons and ions in both the solar wind (e.g. \citet{vstverak2008electron, innocenti2019onset, micera2021role}) and  magnetospheric (e.g. \citet{alexandrova2020situ} ) environment. When plotting electron quantities, one can quickly identify firehose- (bottom right corner) or whistler- (upper right corner) unstable plasma parcels. The area in between the two families of isocontour lines is stable to both instabilities. In this case, we can use this visualization to quickly appreciate the differences between the plots associated with the different clusters. 
In all panels the simulated data points sit in the stability region bounded by the stability thresholds. At initialization, panel a, the data points are well into the stable region. In panel b we see that the majority of data points in cluster 0, inflow cluster, still sit quite close to initialization values at $\Omega_{ci}t= 320$, as expected from regions of space barely influenced by the development of the plasmoid instability. A small number of points (darker color points) has moved to larger parallel beta, still within the stable region. In all the clusters associated with plasmoids the majority of points (lighter color points) have moved to larger parallel electron beta with respect to initialization, due to the larger electron pressure within the plasmoids with respect to the inflow region (see Fig. \ref{sec:PIC}). Comparing panel c, d, e and f, one notices immediately the difference between panel d (cluster 2, green, inner plasmoid region) and the others. In the inner plasmoid region the spread of the points around the core of the distribution is quite small with respect to the other clusters, and all points are quite far from instability thresholds. "Walking" from the plasmoid interior towards the inflow region, and crossing from cluster 2, green, panel d, to cluster 4, purple, panel f, to cluster 1, orange, panel c, and finally to cluster 3, red, panel e, one notices that the data distribution progressively moves towards higher electron parallel beta, still within the stable region. At the separatrix cluster, panel e, the data points have spread into the narrow stable region at high parallel beta.

\begin{figure}
	\centering
	\includegraphics[width=1\textwidth]{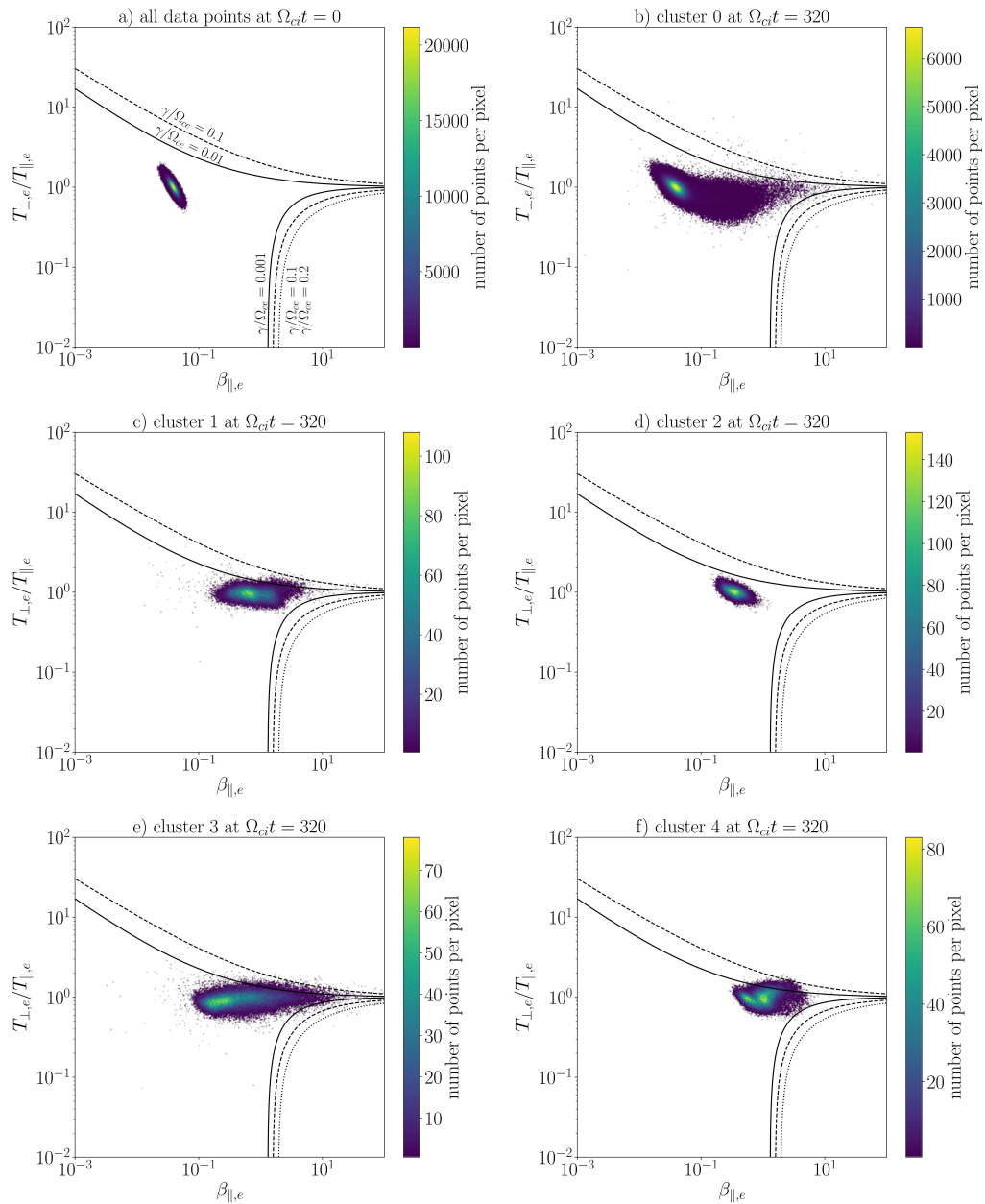}
    \caption{Data point distribution in the $\beta_{\parallel, e}$ vs $T_{\perp,e}/ T_{\parallel,e}$ plane. Panel a: all upper current sheet simulated points at initialization. Panel b to f: points associated with cluster 0, 1, 2, 3, and 4 at $\Omega_{ci}t= 320$. The colors highlight the number of points per pixel. The isocontours of growth rates $\gamma /\Omega_{ce} = 0.001, 0.1, 0.2$ for the resonant electron firehose instability and of growth rates $\gamma /\Omega_{ce} = 0.01, 0.1 $ for the whistler temperature anisotropy instability are depicted in the upper and lower semiquadrant.  }
    \label{fig:Brazil}
\end{figure}

\section{Discussion and conclusion}

In this paper, we have applied an unsupervised clustering method based on SOMs to data points obtained from a fully kinetic simulation of plasmoid instability. The clustering method and the simulation used for the clustering experiment are described in Section \ref{sec:SOM} and Section \ref{sec:PIC}. Pre-processing activities and clustering results obtained with three different types of data scaling are described in Section \ref{sec:res}. We remark on the fundamental role of scalers in determining clustering results. Our dataset includes outlier-poor and outlier-rich features, with the latter being particularly critical in the identification of magnetic reconnection regions. In this situation, a scaler designed to be robust to the presence of outliers, such as the \textit{RobustScaler}, delivers results that match better with our a-priori knowledge of the process under investigation. The data scaled with \textit{RobustScaler} are clustered in an unsupervised fashion into clusters that we identify, a posteriori, as an inflow cluster (cluster 0, blue in Fig.~\ref{fig:scalers:PP}, panel c), an inner plasmoid cluster (cluster 2, green), a separatrix cluster (cluster 3, red), a plasmoid merging cluster (cluster 4, purple) and a further cluster for regions intermediate between the separatrices and the area where plasmoid merging signatures are stronger (cluster 1, orange).\\ In Section \ref{sec:posteriori} we further examine our classification results in light of the analysis tools that SOMs provide us with, mainly feature maps, Fig. \ref{fig:featuremap}, and the Unified Distance Matrix (UDM), Fig.~\ref{fig:scalers:UDMmmst}. We recover from this analysis valuable information on the physical processes occurring in each cluster. In our case, where we are already acquainted with the dataset under investigation and with the physical results of interest, this a-posteriori analysis results into an analysis procedure where information is made readily available in an easy-to-grasp 2D representation, thus simplifying further investigation. Furthermore, we see the potential of using feature and UDM maps as a tool for scientific discovery when applied to datasets encoding unknown processes. We consider of particular interest the analysis of UDM patterns illustrated in Fig.~\ref{fig:handpick_highJz} to Fig.~\ref{fig:handpick_cluster_intersection}: we show that darker patterns in the UDM (meaning, nodes "significantly" different from their neighbors) map to small-scale regions of interest, such as regions in the 2D simulated plane where plasmoid merging signatures are particularly strong, or where smaller-scale plasmoids develop in the current sheet in between larger-scale plasmoids. Also here, we see the double potential of the method, as a fast tool for identifying already-known regions of interest in a potentially very large dataset, and as a tool of scientific discovery when applied to an unknown dataset.\\  
This work is a follow up of previous activities, where a similar clustering method had been applied to MHD simulations. We verify here that this clustering technique delivers excellent and insightful results also with fully kinetic simulations, when it is fed data (moments separated by species, parallel and perpendicular pressure terms, electric field "including" non ideal processes, ...) comparable, in temporal and spatial resolution and in the nature of the processes of interest, to observed data from solar wind and magnetospheric missions.\\
Spacecraft observations are thus a natural field of future applicability for this method. Additionally, we envision a role of this and similar clustering methods in model development. An open issue in multi-physics and coupled methods (e.g., the above-mentioned \citet{daldorff2014two, innocenti2015introduction, ashour2015multiscale, lautenbach2018multiphysics}) is how to decide when to switch between numerical methods (e.g. MHD to PIC, or 10-moment-method to Vlasov) or between lower and higher resolution. This choice is often made somehow empirically, e.g. based on the expected location in space of target processes, or on thresholds of specific quantities. We envision the possibility of using SOMs trained on reference simulations to decide where to perform this switch: specific SOM nodes or specific clusters of SOM nodes could be associated to a certain model or resolution. We remark that, while training a SOM is (moderately) time consuming, BMU evaluation is in fact fast enough to be embedded into a run-time code without significant performance degradation. Preliminary and yet unpublished tests confirm that SOMs are robust to being deployed in simulations performed with a different physical model with respect to the one of training: in \citet{HelioPres}, the SOM nodes mapping to diffusion region and separatrices in Vlasov-Maxwell simulations of magnetic reconnection are shown to identify the same areas in a 10-moment simulation performed with the same physical parameters.\\

\section*{Acknowledgments}

We thank Giovanni Lapenta and the ECsim development team at KULeuven for early access to the code, the AIDA WP7 team (Jorge Amaya, Jimmy Reader, Banafsheh Ferdousi and James "Andy" Edmond) for long discussions on the methodology (\url{https://www.aida-space.eu}) and Richard Sydora for useful feedback on the draft. 

\section*{Funding}
This work was funded by the German Science Foundation DFG within the Collaborative Research Center SFB1491. We gratefully acknowledge the Gauss Centre for Supercomputing e.V. (\url{www.gauss-centre.eu}) for providing computing time on the GCS Supercomputer SuperMUC-NG at Leibniz Supercomputing Centre (\url{www.lrz.de}). This work also used the DiRAC Extreme Scaling service at the University of Edinburgh, operated by the Edinburgh Parallel Computing Centre on behalf of the STFC DiRAC HPC Facility (www.dirac.ac.uk). This equipment was funded by BEIS capital funding via STFC capital grant ST/R00238X/1 and STFC DiRAC Operations grant ST/R001006/1. DiRAC is part of the UK National e-Infrastructure.

\section*{Declaration of interest}
The authors report no conflict of interest.

\section*{Data availability statement}
The data that support the findings of this study are openly available in Zenodo 
at \url{https://zenodo.org/record/7463339#.Y6R2GhWZPEb}, reference number 7463339 \cite{Koehne22_Zenodo}. The scripts are available on GitHub at \\
\url{https://github.com/SophiaKoehne/unsupervised-classification}, \cite{Koehne22_GitHub}.\\

\section*{Author ORCID}
S. K\"ohne, \url{https://orcid.org/0000-0002-0163-0456}\\
E. Boella, \url{https://orcid.org/0000-0003-1970-6794}\\
M.E. Innocenti, \url{https://orcid.org/0000-0002-5782-0013}\\

\section*{Author contribution}
M.E.I. performed the simulations. S.K. performed the clustering. All authors contributed to analyzing data and reaching conclusions, and in writing the paper.

\appendix
\section{Robustness tests}
\label{sec:appendix}
In this Section we illustrate several tests run with the objective of assessing the robustness of our clustering procedure.
First, we want to understand how dependent our clustering results are on the SOM hyper-parameters, see Section \ref{sec:SOM}.
In Table \ref{tab:matchingR:all_new} we list several SOM training experiments, run using as input the upper current sheet data scaled with the \textit{RobustScaler}. We change the number of epochs, the initial learning rate $\eta_0$, the initial neighborhood width $\sigma_0$ (calculated as a difference percentage of the largest map side), the random seed used for random weight initialization, the number of SOM nodes $q$. We then list, in the last column, the percentage of points classified in the same k-means cluster as with a 5 epoch-long training, $\eta_0 = 0.5$, $\sigma_0 = 0.2\times\max(x,y)$, random initialization seed. The k-means clustering is done afresh on the newly trained map. \textcolor{black}{Since the cluster number in k-means is assigned arbitrarily, clusters which are assigned the same number in different clustering experiments do not necessarily correspond to similar regions of simulated space. For this reason, before calculating $R$, we reassign cluster numbers, so that the clusters compared during the $R$ calculation indeed match to similar regions of physical space in the simulations. }\\ 
In calculating the matching factor $R$ we exclude \textcolor{black}{the cluster mapping to the} inflow region \textcolor{black}{(cluster 0, blue)}, which stays essentially constant notwithstanding the SOM hyper-parameters, since we are primarily interested in clusters mapping to the plasmoid region. Without excluding the inflow cluster, the matching factors increase significantly in all cases, simply due to the amount of data points falling into the inflow region.

\begin{table}
    \caption{\textcolor{black}{Percentage $R$ of samples classified in the same k-means cluster as with a map trained for five epochs, initial learning rate $\eta_0 = 0.5 $, initial neighborhood radius $\sigma_0 = 0.2\times\max(x,y)$, random initialization seed 42 and number of nodes $q= 71\times93$ when the number of epochs, $\eta_0, \sigma_0$, random initialization seed and $q$ are varied. In all cases, the scaler used is \textit{RobustScaler}. The inflow cluster, cluster 0, is excluded in the calculation of the matching factor, since we are primarily interested in clusters mapping to the plasmoid region. }}
    \centering
    \begin{tabular}{c c c c c c}
    epochs & $\eta_0$ & $\sigma_0 / \%\max(x,y)$ & random init seed & $q$ & Matching factor $R/\%$ \\
    \hline\hline
    5 & 0.3 & 20 & 42 & 71$\times$93 & 93.55\\
    5 & 0.5 & 20 & 42 & 71$\times$93 & 100 \\
    5 & 0.7 & 20 & 42 & 71$\times$93 & 92.67 \\
    5 & 1 & 20 & 42 & 71$\times$93 & 91.99 \\
    \hline\hline
    \textcolor{red}{5} & \textcolor{red}{0.5} & \textcolor{red}{10} & \textcolor{red}{42} & \textcolor{red}{71$\times$93} & \textcolor{red}{85.78}\\
    5 & 0.5 & 20 & 42 & 71$\times$93 & 100\\
    5 & 0.5 & 40 & 42 & 71$\times$93 & 91.75 \\
    5 & 0.5 & 60 & 42 & 71$\times$93 & 87.78 \\
    \hline\hline
    5 & 0.5 & 20 & 42 & 71$\times$93 &  100 \\
    5 & 0.5 & 20 & 0 & 71$\times$93 & 93.6 \\
    5 & 0.5 & 20 & 1 & 71$\times$93 & 94.47 \\
    \hline\hline
    5 & 0.5 & 20 & 42 & 64$\times$83 & 91.48 \\
    5 & 0.5 & 20 & 42 & 71$\times$93 & 100 \\
    5 & 0.5 & 20 & 42 & 78$\times$102 &  93.23\\
    5 & 0.5 & 20 & 42 & 85$\times$110 & 93.58 \\
    5 & 0.5 & 20 & 42 & 91$\times$118 & 93.99 \\
    \textcolor{blue}{5} & \textcolor{blue}{0.5} & \textcolor{blue}{20} & \textcolor{blue}{42} & \textcolor{blue}{101$\times$132} & \textcolor{blue}{94.47}\\
    \hline\hline
    3 & 0.5 & 20 & 42 & 71$\times$93 & 92.13\\
    4 & 0.5 & 20 & 42 & 71$\times$93 & 91.16\\
    5 & 0.5 & 20 & 42 & 71$\times$93 & 100\\
    6 & 0.5 & 20 & 42 & 71$\times$93 & 89.77\\
    7 & 0.5 & 20 & 42 & 71$\times$93 & 88.4\\
    \end{tabular}
    \label{tab:matchingR:all_new}
\end{table}

We notice that the matching factors are very high for all cases in Table \ref{tab:matchingR:all_new}: our classification procedure is significantly stable to the choice of SOM hyper-parameters. The high matching factors can be explained noticing that map nodes may individually change with different hyper-parameters, but not so much as to give rise to significantly different k-means clustering results. The parameter that gives rise to the most significant differences is 
\textcolor{black}{the initial neighborhood radius}.
\textcolor{black}{In deciding which initial neighborhood radius to use in the body of the manuscript, we followed the rule of thumb prescription from \cite{kohonen2014matlab}. We see here that, with smaller neighborhood radius and training on the same number of epochs, the map may have more difficulties reflecting the true structure and topology of the data.}

\begin{figure}
\centering
\includegraphics[width=0.8\textwidth]{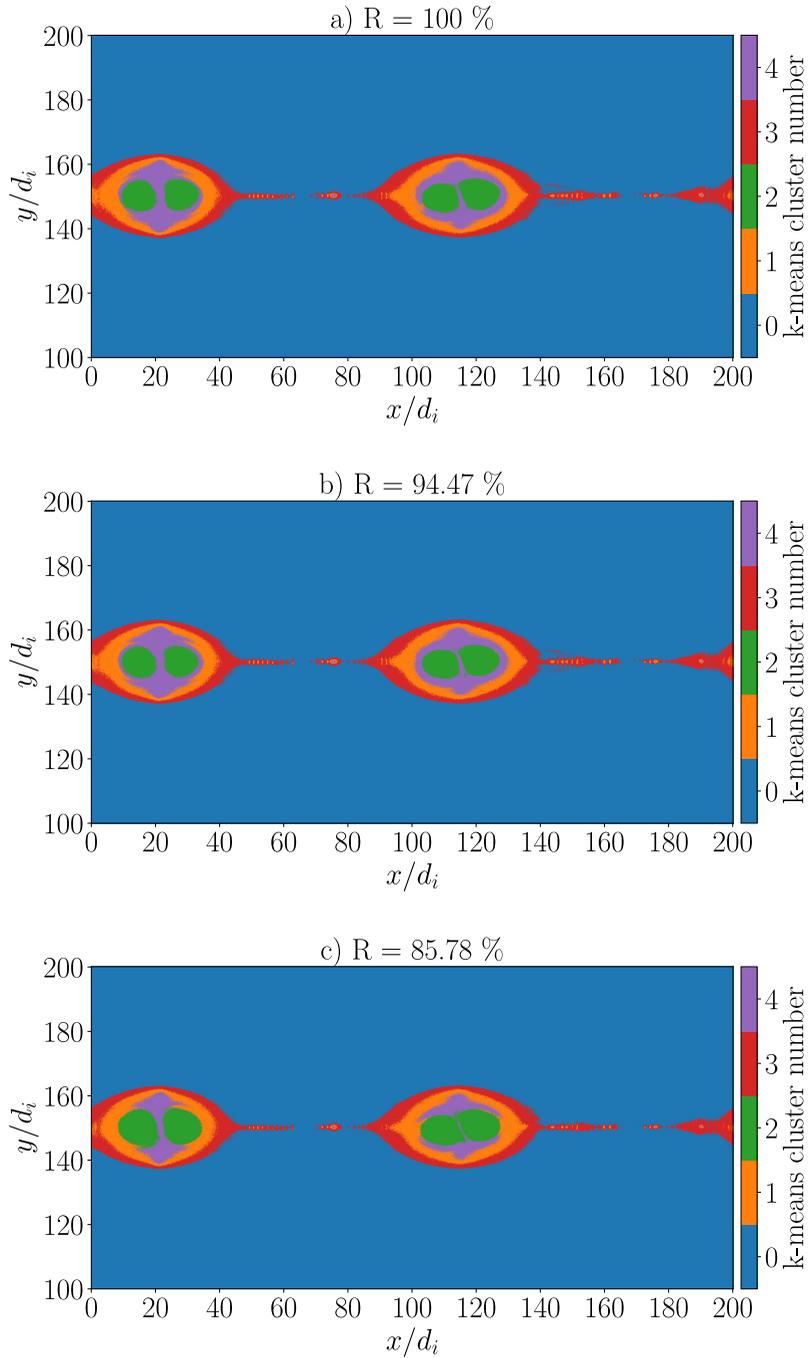}%
\caption{\textcolor{black}{Clustering of the upper current sheet simulated data, with the SOM used as a reference in Table \ref{tab:matchingR:all_new} in panel a and the SOMs trained with the hyper-parameters listed in the lines colored in blue and red in panels b and c.}}
\label{fig:robust:best_worst_R}
\end{figure}

It is interesting to identify which points in the simulation change clusters "more easily". In Fig.~\ref{fig:robust:best_worst_R} we plot the clustered points obtained from the SOM trained with the hyper-parameters listed in red and blue in Table~\ref{tab:matchingR:all_new} \textcolor{black}{compared to the reference SOM in panel a}. We notice that the regions of space which are more affected by the choice of hyper-parameters are those associated with cluster 1, orange and cluster 4, purple, i.e. the intermediate cluster region. This is not a surprise: the two clusters are the most similar and in fact tend to merge when the cluster number is reduced from $k=5$ to $k=4$. Cluster 2, inner plasmoid region, and cluster 3, separatrix cluster, are instead significantly different from the intermediate plasmoid region. \textcolor{black}{The individual matching factors for each cluster in the case with the lowest matching factor $R = 85.78 \%$ are: orange(1): 86.45\%, green(2): 99.4\%, red(3): 91.79\%, purple (4): 64.37\%.} 
\textcolor{black}{We see that results for cluster 0, 3, 2 (the clusters that physics tells us should be clearly distinct) are stable with all hyperparameters, even in the worst case scenario, as we expect from the physics of the problem. Under this point of view, the fact that a small subset of simulated points tend to be assigned, in different clustering experiments, to either one of the two most similar cluster is not considered a cause for concern.   }

\begin{figure}
\centering
\includegraphics[width=0.8\textwidth]{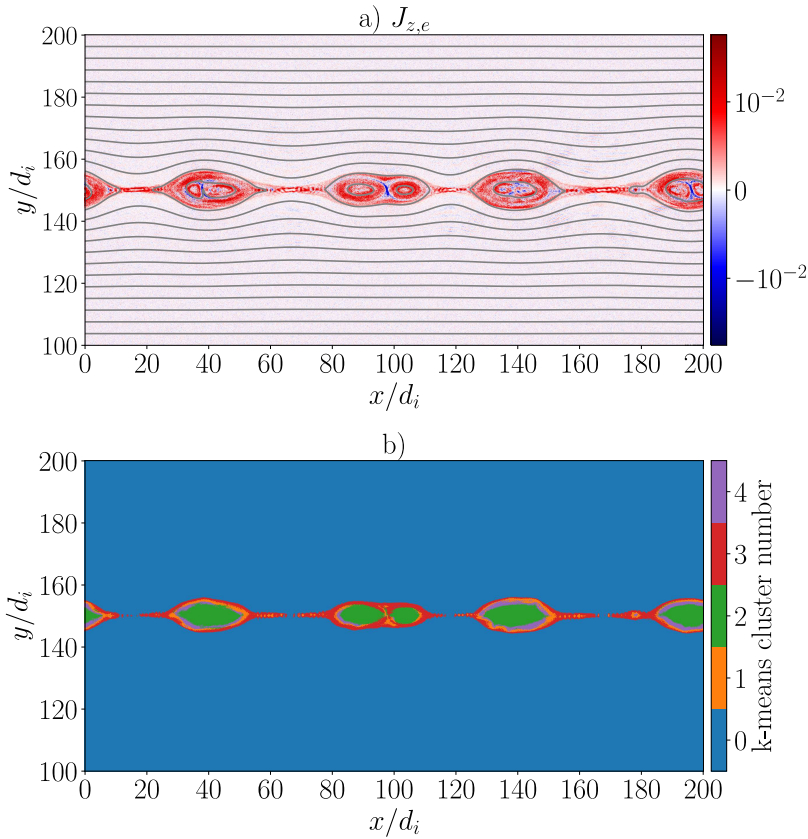}%
\caption{Out-of-plane electron current, $J_{z,e}$, panel a, and clustering of the upper current sheet data points, panel b, at $\Omega_{ci}t=160$. The SOM used for the clustering has been trained at $\Omega_{ce}t= 320$.}
\label{fig:1000:streamline}
\end{figure}

We also want to check the robustness of our clustering procedure to temporal variation in the simulation. To do this, data samples from an earlier time in the simulation, $\Omega_{ci}t= 160$, were classified using the map and k-means centroids described in Section \ref{sec:res} and Section \ref{sec:posteriori}, trained on the data from the timestep $\Omega_{ci}t= 320$ scaled with \textit{RobustScaler}. The BMUs of the data samples from the timestep $\Omega_{ci}t= 160$ were found in that trained map and colored according to the k-means clusters visible in Fig.~\ref{fig:scalers:UDMmmst}, panel e. The out-of-plane electron current and the resulting classification are shown in Fig.~\ref{fig:1000:streamline}, panel a and b.  \\
We immediately notice that, at $\Omega_{ci}t= 160$, the simulation is at a quite different stage. More, smaller plasmoids are present, since the merging process is just at the beginning. This is reflected in the clustering results.\\
In Fig.~\ref{fig:1000:streamline}, panel b, we observe that the inflow and plasmoid region are very well separated. Inside the plasmoid, the inner plasmoid region (cluster 2, green) occupied a larger percentage of the plasmoid area with respect to later times. The cluster associated to plasmoid merging, cluster 4, purple, and cluster 1, orange, have shrunk accordingly, possibly due to a combination of two factors: smaller plasmoid size (the plasmoid size increases after coalescing) and the fact that plasmoid merging is just at the beginning.  

\clearpage
\bibliographystyle{jpp}

\bibliography{lit}

\end{document}